\documentclass[aps,prb,twocolumn,showpacs,amsmath,amssymb,superscriptaddress,10pt]{revtex4-1}

\usepackage[T1]{fontenc}
\usepackage[utf8]{inputenc}
\usepackage[english]{babel}

\usepackage{amssymb}
\usepackage{amsmath}
\usepackage{xcolor}

\usepackage{bm}

\usepackage[pdftex]{graphicx}
\usepackage[colorlinks=true,citecolor=blue]{hyperref}
\usepackage{physics}	
\usepackage{ulem} 
\usepackage{dsfont}
\usepackage[multidot]{grffile}

\usepackage{blindtext}

\def\up{\mathord{\uparrow}}
\def\down{\mathord{\downarrow}}

\begin{document}
\title{Interaction-induced current asymmetries in resonant transport through interacting quantum-dot spin valves revealed by iterative summation of path integrals}

\author{S.~Mundinar}
\author{A.~Hucht}
\author{J.~K\"onig}
\author{S.~Weiss}
\affiliation{Theoretische Physik, Universit\"at Duisburg-Essen and CENIDE, 47048 Duisburg, Germany}

\date{\today}

\begin{abstract}
Resonant tunneling of electrons between two ferromagnets and a quantum dot in the presence of an externally applied magnetic field reveals a strong gate dependence in the linear and nonlinear bias regime. This gate dependence originates from the interplay between Coulomb interactions and spin-dependent hybridization between the quantum dot and the leads. 
To take into account Coulomb interaction strengths of the same order of magnitude as the external magnetic field and the hybridization strength we adopt the numerically exact iterative summation of path integrals (ISPI). 
\end{abstract}

\maketitle

\section{Introduction}\label{sec:intro}

The field of spintronics is under vivid research, where the sphere of interest has broadened to the search for new materials, new concepts for spintronic devices, as well as new functionalities. One of the most fundamental devices of spintronics is the spin valve, which found application e.g. in magnetic sensors, as well as in magnetic random access memory devices (MRAM). 
Spin valves are, e.g., realized in layered heterostructures, which show a strong tunnel magnetoresistance (TMR) effect \cite{Parkin_2004,Ikeda_2008}. 
These devices have strongly profited from the discovery of suitable 2D magnetic materials in recent years \cite{Song_2018, Gong_2017, Huang_2017}. 
Spin valves can also be realized in quantum-dot devices.
These systems allow for a detailed control over single spins. As a result, the impact of spin dynamics on quantum transport properties of such devices and also the manipulation of spins in these systems is of great interest. 

On an experimental level, several effects of quantum-dot spin valve systems have been under investigation. It is possible to control transport of single spins, and therefore the TMR, via a gate voltage. This has been shown for an InAs quantum dot coupled to Ni electrodes \cite{Hamaya_2007_2} as well as for a carbon-nanotube quantum-dot setup \cite{Sahoo_2005}. An external magnetic field has been demonstrated to compensate the exchange field in these experiments, thus effectively enabling a measurement of the lead-induced exchange field. Also measurements of the Kondo effect have been reported for both an InAs and as well as a carbon-nanotube quantum-dot setup \cite{Hamaya_2007_1,Hauptmann_2008}. Again, for a setup including a carbon nanotube as quantum dot the precession of an on-site spin can be harnessed and tuned via a gate voltage \cite{Crisan_2016}, and the magnetoresistance shows a strong hysteretic behavior for temperatures lower than 30 K \cite{Jensen_2005}. 

From a theoretical point of view, the interplay between finite Coulomb interaction, nonequilibrium dynamics and spin-dependent resonant tunneling is challenging. 
The combination of Coulomb interactions and spin-dependent tunneling gives rise to an exchange field.
It yields an interaction-induced spin precession, that - up to first order in the tunnel coupling - depends sensitively on all system parameters including bias voltage and temperature \cite{Braun_2004}. 
Based on different approximation strategies, there exist several sophisticated methods which are appropriate to tackle the interacting, spin-dependent problem in a nonequilibrium situation. 
By treating electron-electron interactions in Hartree-Fock approximation, the TMR and the accumulated spin were examined for a quantum-dot spin valve in a noncollinear setup \cite{Rudzinski_2005}. In the high-temperature limit, several peaks in the quantum-dot spin valve's conductance were discussed using a hierarchical quantum master equation approach \cite{Wenderoth_2016}. Within a weak-tunneling approximation in a wide range of bias and gate voltages, characteristics of the exchange field for a quantum-dot spin valve have been discussed \cite{Braun_2004, Koenig_2003}. Also a multi-level island coupled to ferromagnetic leads \cite{Lindebaum_2011, Lindebaum_2012} has been treated on the same footing. It has been shown that the exchange field could be probed via a third, superconducting lead \cite{Sothmann_2010_1} as well as via a spin resonance, that appears when the exchange field is perpendicular to one of the leads polarizations \cite{Hell_2015}. The coherent dynamics of a quantum-dot spin valve have been shown to be distillable from full counting statistics \cite{Stegmann_2018}. Recently, tuning gate voltages on a quantum dot spin valve has been suggested to switch magnetizations of the attached leads \cite{Gergs_2018}.

A strong influence of resonant tunneling on the TMR \cite{Mundinar_2019} shows that it is essential to include all orders of the tunneling for a complete picture of the electronic transport through small quantum-dot spin valves. For the regime of linear response NRG approaches work well \cite{Weymann_2011,Simon_2007, Sindel_2007}. Using a DMRG approach, the local density of states as well as the TMR have been calculated \cite{Gazza_2006}. The Kondo problem for a quantum-dot spin valve was tackled via an equations-of-motion method \cite{Swirkowicz_2008}. For the sake of completeness, we note that quantum-dot systems attached to normal metal leads are also studied by means of quantum Monte Carlo simulations \cite{Lichtenstein_2019}, allowing, e.g. the study of the time-dependent current through these systems as well as the Green's functions and occupation numbers \cite{Antipov_2017}. The short-to-intermediate time limit for, e.g. the tunneling current could also be obtained by means of a generalized iterative influence functional \cite{Simine_2013}. 

In this work, we investigate the spin dynamics in resonant transport through a quantum-dot spin valve. We focus on the interplay between Coulomb interaction, a noncollinear setup of the leads' magnetizations and a local Zeeman field acting on the quantum dot in arbitrary direction. We find a strong asymmetry in the current as a function of the gate voltage. This effect is attributed to the noncollinearity of the setup, the Zeeman field as well as the Coulomb interaction, which gives rise to an exchange field. Numerically exact iterative path integral summations (ISPI) are carried out, where the scheme is adopted for the spin-dependent case here. 
As a further development of our method, local observables such as the mean occupation of the dot and spin expectation values are calculated within the scheme. The technique is formulated on the Keldysh contour and builds upon systematic truncation of lead induced correlations. It has already been used to describe the current through the Anderson model \cite{Weiss_2008,Weiss_2013}. Current and TMR for transport through a quantum-dot spin valve in collinear setup were also discussed in an earlier work in Ref.~\onlinecite{Mundinar_2019}. The ISPI method excels in the study of the stationary limit {\color{blue}while}
all energy scales of the system are of the same order of magnitude. In this regime, resonant tunneling leads to a finite peak broadening. In fact, this peak broadening as well as interaction-induced effects are both well resolved within the ISPI scheme, which therefore is a useful tool when discussing interacting quantum-dot spin valves.

Our article is structured as follows. In Sec.~\ref{sec:model} we introduce the Hamiltonian of our system. In  Sec.~\ref{sec:method}, we outline how to obtain expectation values for spin  and occupation number of the quantum dot by calculating  the  coherent state path-integral on the Keldysh contour in  the presence of  suitable source terms. Our results are presented in Sec.~\ref{sec:results}, outlining the convergence procedure in order to obtain numerically exact results. A benchmark for the noninteracting system is presented in Sec.~\ref{ssec:benchmark}.  We present interaction-induced current asymmetries in Sec. \ref{ssec:asymmetry}. 
To complete the picture, we study the dependence of the electronic transport on the angle between the leads' magnetizations  in Sec.~\ref{ssec:angle}.

\section{Model}\label{sec:model}
\begin{figure}[t]
\centering
\includegraphics[width=\columnwidth]{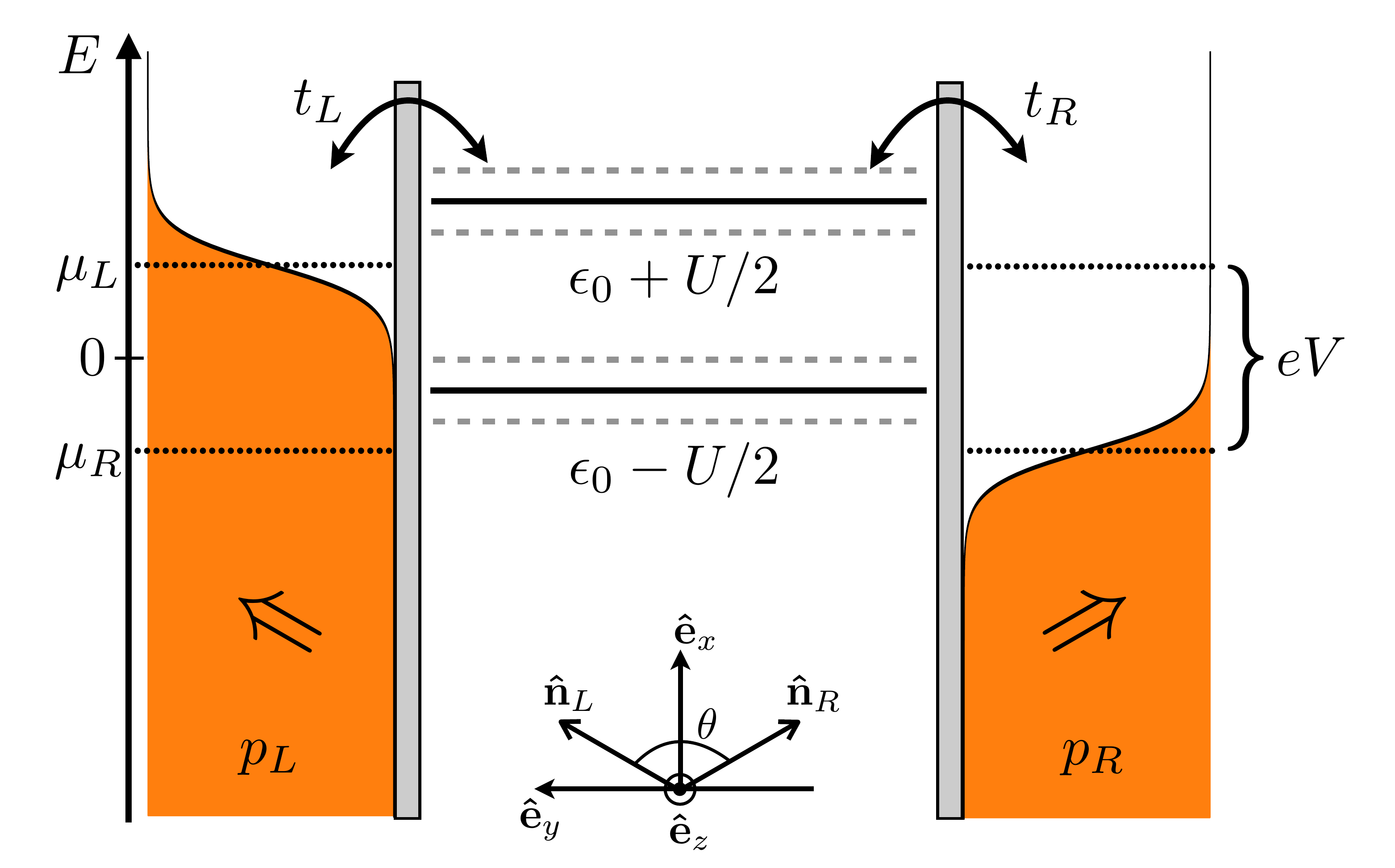}
\caption{A single-level quantum dot coupled to ferromagnetic leads $\alpha=L/R$ with magnetization axes $\vu{n}_\alpha$. The polarization strength $p=p_\alpha$ induces spin dependent hybridizations $\Gamma_{\alpha\tau} = 2\pi \rho_{\alpha \tau} t^2_\alpha$. The Coulomb interaction is denoted $U$ and the single particle energy is $\epsilon_0$. A bias voltage $eV=\mu_L-\mu_R$  drops across the quantum dot. In the presence of an additional Zeeman field, the spin degeneracy is lifted (sketched by dashed lines).}
\label{fig:system}
\end{figure}

We study an interacting quantum-dot spin valve, which comprises a quantum dot coupled to two ferromagnetic leads, see Fig. \ref{fig:system} for a sketch of the system. Throughout this work we set $\hbar = 1$. The system is described by a Hamiltonian consisting of three parts
\begin{equation}
	\mathcal{H} = \mathcal{H}_{\text{dot}} + \mathcal{H}_{\text{leads}} + \mathcal{H}_{\text{T}}.	
\end{equation}
Electrons in the ferromagnetic leads $\alpha = L/R=\pm 1$ are described as a free electron gas
\begin{equation}\label{eq:Hleads}
	\mathcal{H}_{\text{leads}} = \sum_{\alpha \vb{k} \tau} c^\dagger_{\alpha \vb{k} \tau} (\epsilon_{\vb{k} \tau} - \mu_\alpha) c_{\alpha \vb{k} \tau}, 
\end{equation}
where $\epsilon_{\vb{k} \tau}$ is the single-electron energy for spin projection $\tau=\pm$, and $\mu_\alpha$ is the chemical potential. Since we choose the quantization axes of each lead along its respective magnetization $\vu{n}_\alpha$, the spin index $\tau$ describes majority ($+$) and minority ($-$) spins. Assuming the density of states $\rho_{\alpha \tau} = \sum_{\vb{k}} \delta(\omega - (\epsilon_{\vb{k} \tau} - \mu_\alpha))$ constant and spin dependent, the asymmetry between majority and minority spins is characterized by the polarization $p_\alpha = \frac{\rho_{\alpha +} - \rho_{\alpha -}}{\rho_{\alpha +} + \rho_{\alpha -}}$. Possible values are $0 \leq p_\alpha \leq 1$, where $p_\alpha=0$ describes a nonmagnetic electrode and $p_\alpha=1$ a halfmetallic one. As a common coordinate system for dot and leads we choose $\vu{e}_x = \frac{\vu{n}_R+\vu{n}_L}{\abs{\vu{n}_R+\vu{n}_L}}$, $\vu{e}_y = \frac{\vu{n}_R-\vu{n}_L}{\abs{\vu{n}_R-\vu{n}_L}}$ and $\vu{e}_z = \vu{e}_x \cross \vu{e}_y$. As a consequence, both $\vu{n}_\alpha$ belong to the $x-y$-plane and enclose the angle $\theta$. 

For the quantum dot we assume a single, spin-degenerate orbital of energy $E_0$. Electrons on the dot are subject to a local magnetic field $\vb{B}$ (in units of $g \mu_B$) and the Coulomb interaction $U$, if the orbital is doubly occupied. Hence, the dot Hamiltonian takes the form
\begin{equation}
	\mathcal{H}_\text{dot} = \sum_{\sigma \sigma'} d^\dagger_\sigma [E_0 \delta_{\sigma \sigma'} + ({\vb B} \cdot {\vb S})_{\sigma \sigma'} ] d_{\sigma'} + U d^\dagger_{\up} d^\dagger_{\down} d_{\down} d_{\up},
\end{equation}
where $\vb{S} = \frac{1}{2} \bm{\sigma}$, while $\bm{\sigma} = (\sigma_x, \sigma_y, \sigma_z)$ is the vector of Pauli matrices. The quantization axis of the dot is chosen along the $z$-axis of the coordinate system defined before. Thus the spin index of the dot $\sigma=\up,\down$ denotes the spin projection along the $z$-axis. For later convenience we make use of the operator identity $2d^\dagger_\uparrow d^\dagger_\downarrow d_\downarrow d_\uparrow = d^\dagger_\uparrow d_\uparrow + d^\dagger_\downarrow d_\downarrow - (d^\dagger_\uparrow d_\uparrow - d^\dagger_\downarrow d_\downarrow)^2$ and absorb the quadratic parts into the free Hamiltonian, leading to 
\begin{equation}\label{eq:Hdot}
	\begin{split}
	\mathcal{H}_\text{dot} & = \sum_{\sigma \sigma'} \bigg[d^\dagger_\sigma [\epsilon_0 \delta_{\sigma \sigma'} + ({\vb B}\cdot {\vb S})_{\sigma \sigma'} ] d_{\sigma'} \\
	& - \frac{U}{2} \left(d^\dagger_\sigma (\sigma_z)_{\sigma \sigma'} d_{\sigma'} \right)^2\bigg],
	\end{split}
\end{equation}
with $\epsilon_0 = E_0 + U/2$ being tunable via a gate voltage.

Coupling the leads to the quantum dot is achieved via the tunneling Hamiltonian 
\begin{equation}\label{eq:Htun}
	\mathcal{H}_\text{T} = \sum_{\alpha \vb{k} \tau \sigma} c^\dagger_{\alpha \vb{k}\tau} Y_{\alpha, \tau \sigma} d_\sigma + \text{H.c.},
\end{equation}
where we exploited the explicit $SU(2)$ rotation to the common underlying coordinate system for the symmetric setup. Matrix elements $Y_{\alpha, \tau \sigma}$ reflect spin-dependent tunneling and are the elements of \cite{Weiss_2015}
\begin{equation}\label{eq:SpinDepTun}
Y_\alpha = t_{\alpha} e^{\frac{i\pi}{4}\sigma_y}e^{\alpha \frac{i\theta}{4} \sigma_z} = \frac{t_\alpha}{\sqrt{2}} \left( \begin{array}{cc} e^{i\alpha \theta/4} & e^{-i \alpha \theta/4} \\ -e^{i \alpha \theta/4} & e^{-i\alpha \theta/4} \end{array} \right),
\end{equation}
with $t_\alpha \in \mathbb{R}$ being the tunneling matrix element. Spin-dependent tunneling manifests itself in spin-dependent hybridization strengths, given by $\Gamma_{\alpha \tau} = 2 \pi \abs{t_\alpha}^2 \rho_{\alpha \tau}$. For later use, it is convenient to define the average hybridization strength for lead $\alpha$, which is $\Gamma_\alpha = (\Gamma_{\alpha+} + \Gamma_{\alpha-})/2$. In what follows, we are going to assume a symmetric coupling to the left and right lead and define $\Gamma = \Gamma_L = \Gamma_R$.

\section{Method}\label{sec:method}
Observables of interest are $\langle \mathcal{S}_i\rangle, i=x,y,z$, i.e. different spin-projection expectation values as well as the quantum dot's occupation $\langle \mathcal{N}\rangle$ and the tunneling current $I$ at measurement time $t_m$. Nonequilibrium properties are taken into account within a functional integral formulation on the Keldysh contour \cite{Kamenev, Negele_Orland, Mundinar_2019}. Monitoring these observables allows to obtain a complete picture of the spin dynamics of the system. Within the construction of the functional integral, we will consider source terms that allow to calculate observables $O$ which are given in terms of bilinear products of dot operators, as $O = O(d^\dagger_\sigma d_\sigma, d^\dagger_{\bar \sigma} d_\sigma, d^\dagger_\sigma d_{\bar \sigma})$, which holds for $\langle \mathcal{S}_i\rangle$ and $\langle \mathcal{N}\rangle$, (how to obtain the current is described in Ref.~\onlinecite{Mundinar_2019}). Expectation values may then be computed as
\begin{equation} \label{eq:GenExpVal}
	\expval{O} = -i \fdv{\eta} \ln Z[\eta] \Big|_{\eta=0}.
\end{equation}
Here $Z[\eta]$ is the \textit{Keldysh generating functional} of our system, while $\eta$ is a real-valued parameter. Note that within the Keldysh formalism, propagation in time extends in forward as well as backward direction, usually encoded into a propagation along a closed time contour $C$. As a consequence, the definition $\tilde t = (t,\nu)$, with physical time $t$ and Keldysh branch index $\nu=\pm$, where ($\pm$) represent upper/lower branch, turns out to be useful. 
We evaluate the Keldysh generating functional in the basis of fermionic coherent-states $\bar \psi_{0 \sigma,\alpha \vb{k} \tau}(\tilde t)$, $\psi_{0\sigma,\alpha \mathbf{k} \tau}(\tilde t)$, which are defined by the eigenvalue equations for dot degrees of freedom
\begin{equation}
	\begin{split}
	d_\sigma\,(\tilde t\,) \ket{\Psi\,(\tilde t\,)} = \psi_{0 \sigma}\,(\tilde t\,) \ket{\Psi\,(\tilde t\,)} \\
	\bra{\Psi\,(\tilde t\,)} d^\dagger_\sigma\,(\tilde t\,) = \bra{\Psi\,(\tilde t\,)} \bar \psi_{0 \sigma}\,(\tilde t\,), 
	\end{split}
\end{equation}
and for the leads accordingly. 
As usual, the generating functional within the Keldysh formalism takes the form
\begin{equation} \label{eq:GenFunctionalStart}
	Z[\eta] = \tr e^{i \left( S + \eta O(\tilde t_m)\right)},
\end{equation}
where $S$ is the action of the system \cite{Kamenev,Mundinar_2019, Weiss_2013}, comprising of contributions from dot, leads and tunneling, while $O(\tilde{t}_m)=O[\bar \psi_{0\sigma}(\tilde t_m),\psi_{0\sigma}(\tilde t_m)]$ is the \textit{source term}, which allows to calculate observables at measurement time $t_m$. 
 Performing the trace in Eq.~\eqref{eq:GenFunctionalStart} does not pose any challenge on quadratic terms from leads and tunneling of the Hamiltonian. In order to tackle the quartic interaction term, we first discretize the Keldysh contour into $2N$ time slices of length $\delta_t$ and perform a Hubbard-Stratonovich (HS) transformation on each of these slices \cite{Hubbard_1959, Hirsch_1983}
\begin{equation}\label{eq:HStrafo}
	\begin{split}
	& \exp \left\{ -\nu \frac{i \delta_t U}{2} \left( \sum_{\sigma \sigma'} \bar \psi_{0\sigma}^\nu (\sigma_z)_{\sigma \sigma'} \psi_{0\sigma'}^\nu \right)^2 \right\}  \\ 
	& = \frac{1}{2} \sum_{s=\pm 1} \exp \left\{ - s \lambda_\nu \sum_{\sigma \sigma'} \bar \psi_{0\sigma}^\nu (\sigma_z)_{\sigma \sigma'} \psi_{0\sigma'}^\nu \right\},
	\end{split}
\end{equation}
with $s=\pm 1$ being an Ising-like degree of freedom, and the HS  parameter $\lambda_\nu$ is determined uniquely for $0\leq U < \pi/\delta_t$, see Refs.~\onlinecite{Mundinar_2019, Weiss_2013}. 
As a consequence, it is possible to trace over the dot degrees of freedom within Eq.~\eqref{eq:GenFunctionalStart}, which yields
\begin{equation} \label{eq:discreteGenFunc}
	Z[\eta] = \sum_{\{ \vb{s} \}} \det \left[ \Delta^{-1}[\eta] - \Sigma_C(\vb{s})\right],
\end{equation}
with the vector $\vb{s} = (s^+_1, s^-_1, s^+_2, s^-_2,\ldots, s^+_N, s^-_N)$, hence the sum includes all $2^{2N}$ possible spin configurations along the discretized Keldysh contour where each HS spin couples to the $z$-spin component, see Eq.~\eqref{eq:HStrafo}. We identify the inverse time-discrete Green's function of the noninteracting spin-valve $\Delta^{-1}[\eta] = \Delta_0^{-1} - \Sigma_{\text{T}} + \eta O(t_m)$ in the presence of the source term and the tunneling self energy $\Sigma_T$. Note that only the charging self energy $\Sigma_C(\vb{s})$ depends on all Ising spins $\vb{s}$. 
The matrices have dimensions $4N\times 4N$ due to the Trotter slicing and the spin degree of freedom. 

To specify their matrix elements, we make use of multi-indices $a =(l,\nu,\sigma)$, where the Trotter index $l = 1, . . . , N$ labels the time slice, the Keldysh index $\nu = \pm$ distinguishes the upper from the lower Keldysh contour, and $\sigma =\pm$ labels the spin. The charging self energy $\Sigma_C(\vb{s})$ is time-local and therefore a diagonal matrix
\begin{equation}\label{eq:ChargingSE}
	(\Sigma_C(\vb{s}))_{a a'} = \delta_{aa'}\lambda_\nu \sigma s_l^\nu.
\end{equation}
In the absence of the source term, the Green's function of the dot in the presence of tunneling is given as
\begin{widetext}
\begin{equation}\label{eq:gom}
\left(\Delta^{-1}[0]\right)_{aa'} =  \int \frac{d\omega}{2\pi} e^{-i\omega (l-l')\delta_t} \left[ \nu[(\omega - \epsilon_0)\sigma_0 -\vb{B}\vdot \vb{S}]_{\sigma\sigma'} \delta_{\nu \nu'} \delta_{\sigma\sigma'} 
-\frac{i}{2} \sum_\alpha \left(\check{\Gamma}_{\alpha}\right)_{\sigma\sigma'} \left( F_\alpha (\omega) \right)_{\nu\nu'} \right].
\end{equation}
\end{widetext}
Here the hybridization between dot and lead $\alpha$ is described by the matrix 
\begin{equation}
\check{\Gamma}_{\alpha}=\left(\begin{array}{cc}
	\Gamma_\alpha & e^{-\alpha i \theta/2} p\Gamma_\alpha \\
	e^{\alpha i \theta/2} p\Gamma_\alpha & \Gamma_\alpha \end{array}\right),
\end{equation} 
in spin space which takes into account arbitrary angles $\theta$ between the magnetizations of left and right lead. Furthermore,
$\left( F_\alpha (\omega) \right)_{\nu\nu'}$ are the matrix elements of the $2\times 2$ Keldysh matrix 
\begin{equation}
F_\alpha(\omega) =\left(\begin{array}{cc}2f_\alpha(\omega)-1&-2f_\alpha(\omega)\\ -2f_\alpha(\omega)+2&2f_\alpha(\omega)-1\end{array}\right),
\end{equation}
where the Fermi function $f_\alpha(\omega)=\left[\exp(\beta(\omega-\mu_\alpha))+1\right]^{-1}$ describes the equilibrium occupation distribution of lead $\alpha$.

For the source term $\eta O(t_m)$ in Eq.~\eqref{eq:discreteGenFunc} we outline the functional form for calculating the occupation number operator $\mathcal{N}$ and the operators of the spin projections along the three directions $\mathcal{S}_{x,y,z}$. Using the definition for the lesser Green's function 
\begin{align}
	G^{<}_{\sigma \sigma'}(t,t') &\equiv - i \expval{ T_C \left[d_\sigma(t,+) d_{\sigma'}^\dagger(t',-) \right]}\nonumber\\
	&= i \expval{ d^\dagger_{\sigma'}(t',-) d_\sigma(t,+)},
\end{align}
with  $T_C$ being the contour ordering operator. We find for the observables of interest
\begin{equation} \label{eq:expecValues}	
	\begin{split}
	\expval{\mathcal{N}} (t) = & \tr\left(G^{<}(t,t)\right), \\
	\expval{\mathcal{S}_{x,y,z}} (t) =  & \frac{1}{2} \tr\left(G^{<}(t,t) \cdot \sigma_{x,y,z}\right).
	\end{split}
\end{equation}
Since the operators act on different branches of the Keldysh contour, the limit of equal time arguments $t=t'$ is well defined.
In particular for the occupation number we have the following expression
\begin{equation} \label{eq:SourceTerm}
	\eta \mathcal{N}_{aa'} = \eta \delta_{lm}\delta_{l'm}\delta_{\nu -}\delta_{\nu'+}\delta_{\sigma\sigma'},
\end{equation}
which is a matrix containing two nonzero elements at Trotter slice $m=t_m/\delta_t$. Note that similar expressions are obtained for $\mathcal{S}_{x,y,z}$. For numerical convenience we use the modified generating functional $\tilde{Z}[\eta]\equiv (\det \Delta[0]) Z[\eta] $ rather than Eq.~\eqref{eq:discreteGenFunc} to obtain our results \cite{Weiss_2013,Mundinar_2019}.

Since it is known that lead-induced correlations, and therefore the Green's function of the noninteracting system, i.e. $\Delta[0]_{aa'}$, decay exponentially with increasing $(l,l')$ at finite temperature and/or bias voltage, we truncate those correlation. In turn, the Keldysh generating functional might then be calculated by the numerically exact scheme of iterative summation of path integrals \cite{Mundinar_2019,Weiss_2013,Weiss_2008}, whose main building blocks are outlined in App.~\ref{app:ISPI}.

\section{Results}\label{sec:results}
In what follows, we present results for the symmetric case. This means that the lead polarizations are chosen such that $p_L = p_R = p$, whereas hybridization strengths are $\Gamma_L = \Gamma_R = \Gamma$. The angle $\theta$ between the leads' magnetizations is $\theta_L=-\theta_R=\theta/2$, cf. Eq.~\eqref{eq:SpinDepTun} and the bias voltage is defined by $\mu_L = -\mu_R = eV/2$.  Since we are especially interested in resonant transport and spin effects, $k_BT \ll \Gamma$ in the following. It has been shown before in Ref.~\onlinecite{Mundinar_2019} that a weak-coupling theory does not apply in this regime, while the ISPI method is able to give reliable predictions, once convergence is reached.

\begin{figure}[t]
\centering
\includegraphics[width=\columnwidth]{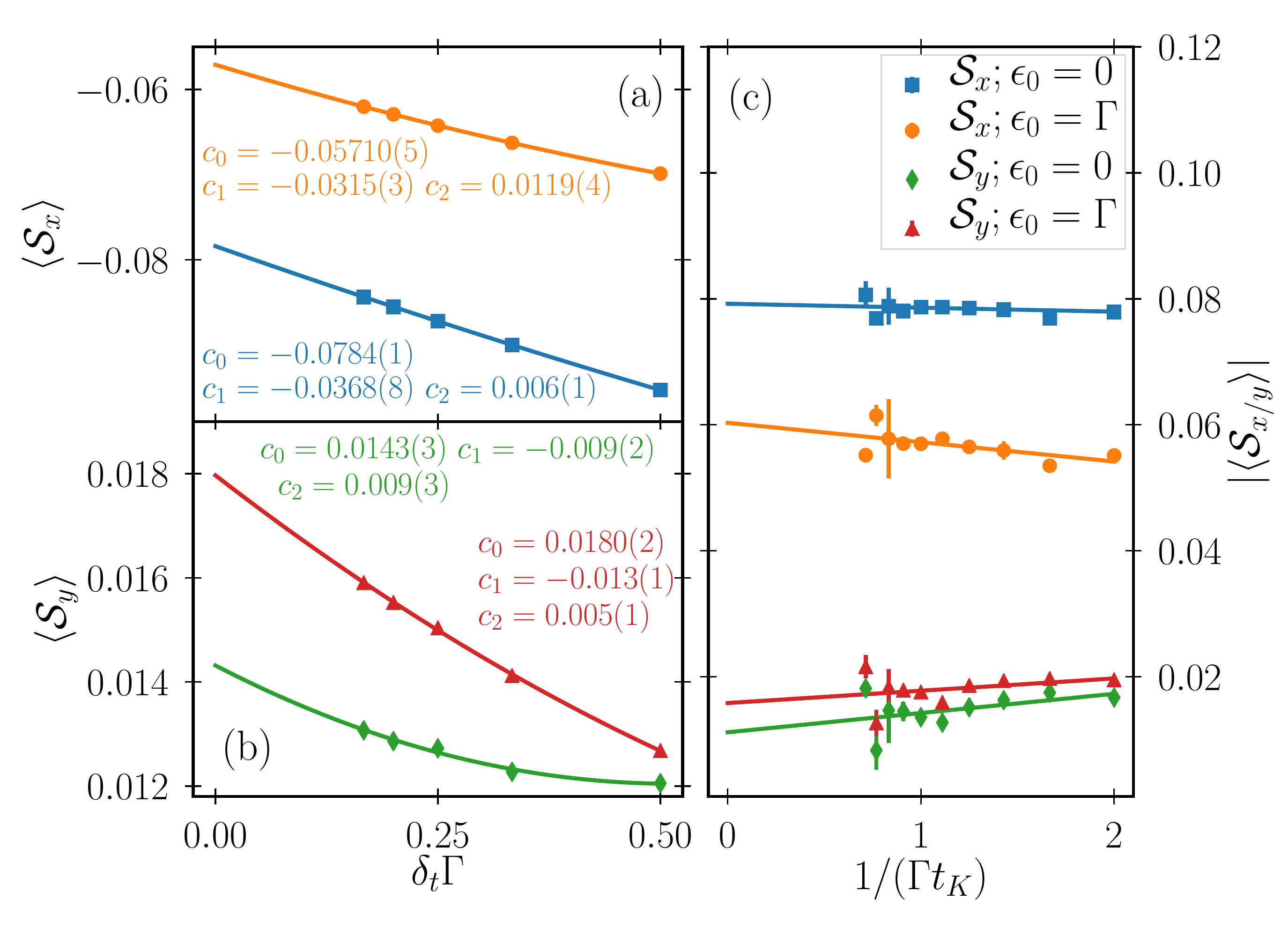}
\caption{Extrapolation procedure for $x$- and $y$-component of spin expectation values. In panel (a) and (b) the elimination of the Trotter error for correlation time $\Gamma t_K=1.0$ is shown as fitting data for different time increments $\delta_t$. (c) Example of the linear regression for the absolute values $|\mathcal{S}_{x,y}|$ as function of $1/(\Gamma t_K)$ after   elimination of the Trotter error. Solid lines represent linear fits to the data. Other parameters are $k_BT=0.2\Gamma$, $U=0.5\Gamma$, $eV=0.1\Gamma$, $B_y=0.2\Gamma$, $p=0.5$, $\theta=\pi/2$.}
\label{fig:Convergence}
\end{figure}

When applying the ISPI scheme, two systematic errors are accumulated, that is (i) the finite time increment $\delta_t$ and (ii) the truncation of lead-induced correlation.
To eliminate both errors, we (i) compute the observable of interest at a fixed correlation time $t_K$ and different values of $\delta_t$. This gives a set of realizations for every observable $O(t_K,\delta_{t,l}), l=2,\dots,7$ which we fit against the polynomial expression $O(t_K,\delta_t\to 0)=\lim_{\delta_t\to 0}\sum_{i=0}^n c_i \delta^i_t$. This means that $O(t_K)$ is given by the constant $c_0$ which inherits a natural deviation due to the (non)linear fitting procedure. In practice we have chosen $\Gamma \delta_t \leq 0.5$ which allows to resolve all physical properties for finite bias voltage and temperature. Note that for current expectation values the fitting function is linear in $\delta_t$, whereas $\langle \mathcal{N}\rangle$ and $\langle \mathcal{S}_i\rangle$ have been fitted with quadratic order to sufficient accuracy for all gate voltages $-2\Gamma < \epsilon_0 < 2\Gamma$. Examples are shown in Fig.~\ref{fig:Convergence} (a), (b) and previously in Refs.~\onlinecite{Mundinar_2019,Weiss_2013}. We repeat the $\delta_t\to 0$ fitting for $0.5\leq t_K\Gamma\leq 2.5$ and if possible extrapolate for $1/(\Gamma t_K) \to 0$, which eliminates the truncation error as well, see Fig.~\ref{fig:Convergence} (c). 
Hence, all presented results are numerically exact, error estimates are given by the standard deviations of the subsequent extrapolations. We are confident that convergence is ensured for small up to intermediate Coulomb interactions of $U\leq2\Gamma$ for the spin-valve quantum dot. Observables reflect their stationary values for $\Gamma t_m>10$ such that  possible transient behavior is absent from the shown data.

\subsection{Benchmark: Noninteracting Case}\label{ssec:benchmark}
\begin{figure}[t]
\includegraphics[width=\columnwidth]{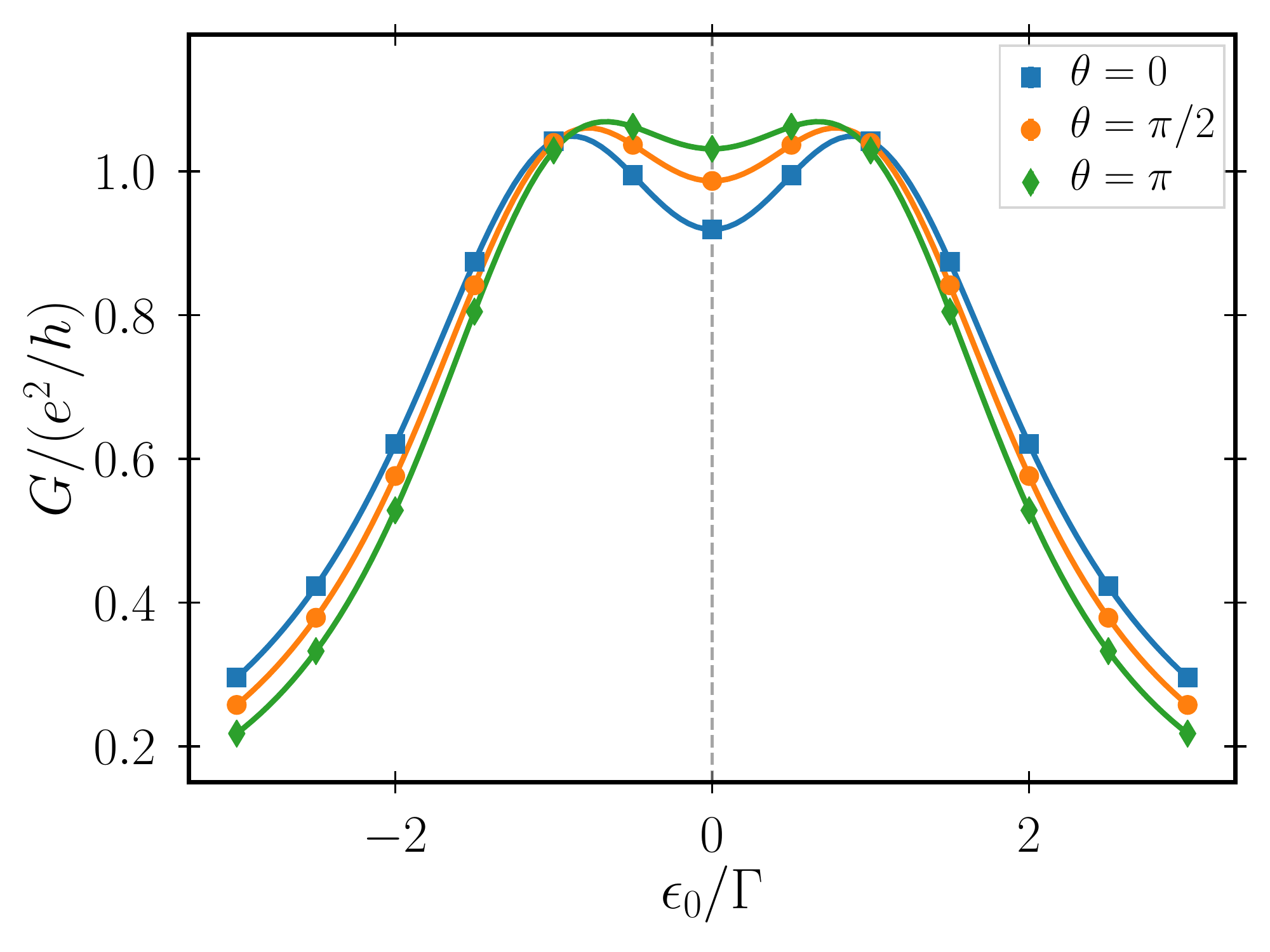}
\caption{Linear conductance through the noninteracting spin-valve as a function of gate voltage $\epsilon_0$ for (non)collinear lead magnetizations. An external Zeeman field of strength $B_z=2\Gamma$ is applied in $x$-direction. Other parameters are $k_BT=0.2\Gamma$, $eV=0.1\Gamma$, $p=0.5$.}
\label{fig:BenchCurrent}
\end{figure}

\begin{figure}[t]
\includegraphics[width=\columnwidth]{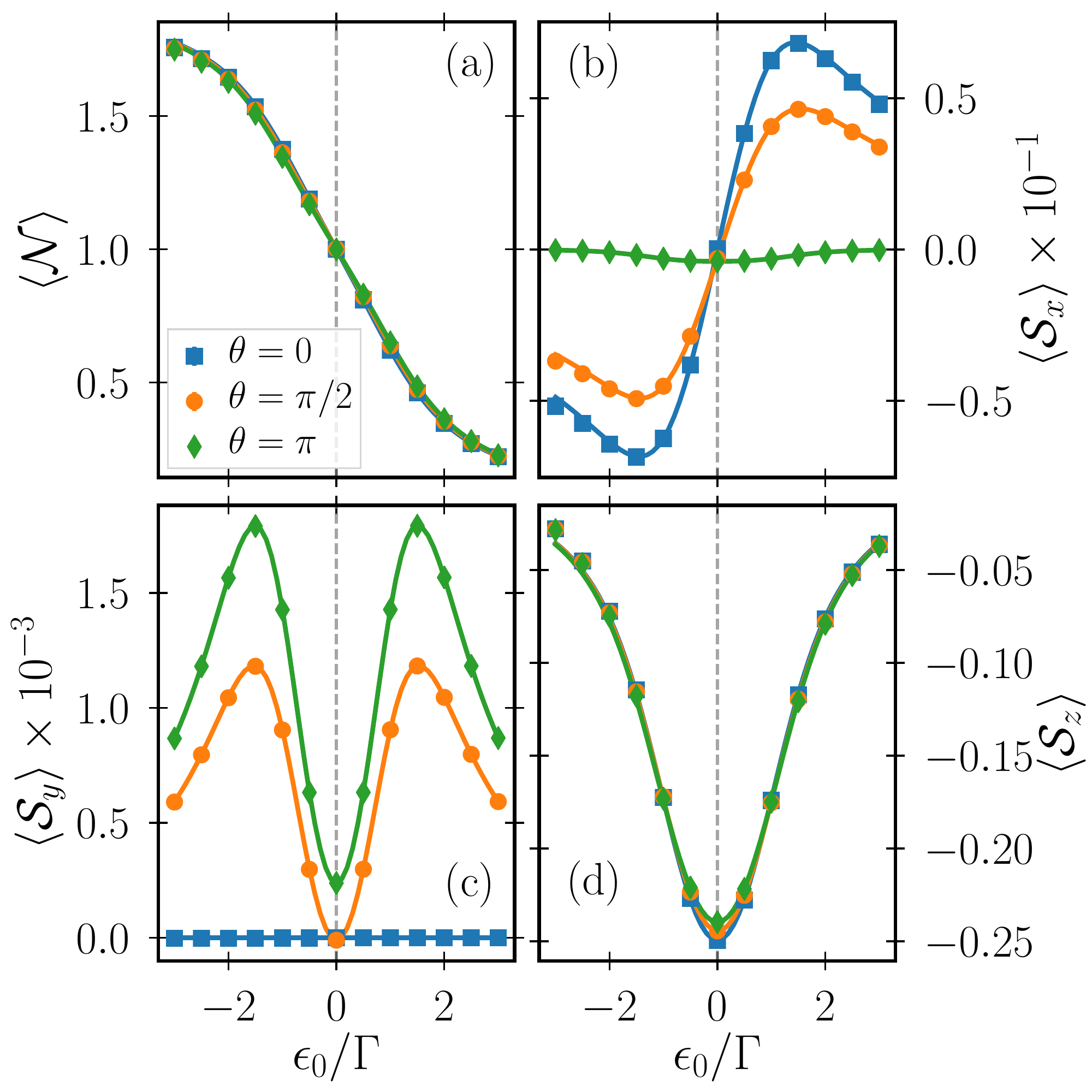}
\caption{The four local observables $\langle \mathcal{N}\rangle$, $\langle \mathcal{S}_{x,y,z}\rangle$, for the noninteracting case as a function of $\epsilon_0$. Parameters are chosen as in Fig.~\ref{fig:BenchCurrent}}.
\label{fig:BenchLocal}
\end{figure}

As a benchmark, we discuss the noninteracting quantum-dot spin valve in the presence of an external Zeeman field. In the absence of interactions, the tunneling current as well as  mean occupation number and spin expectation values are calculated analytically. For the current, a  Meir-Wingreen formula \cite{Jauho_1994} in terms of nonequilibrium Green's functions applies also for the quantum-dot spin valve. The explicit formulae for the respective spin expectation values are given as examples in App.~\ref{app:Analytical}.

In Fig.~\ref{fig:BenchCurrent} the linear conductance in the presence of an external magnetic field $B_z=2\Gamma$ as a function of gate voltage $\epsilon_0$ is shown for three different setups of lead magnetizations, namely parallel ($\theta=0$), orthogonal ($\theta=\pi/2$) and antiparallel ($\theta=\pi$). Further parameters are $k_BT=0.2\Gamma$, $eV=0.1\Gamma$ and $p=0.5$. We compare the analytic results (solid lines) with ISPI data (symbols). Note that ISPI error bars are of the order of the symbol size. 
Resonant transport through a noninteracting single level results in a Lorentzian conductance profile as a function of the gate voltage $\epsilon_0$, with maximum at $\epsilon_0=0$ and width determined by the tunneling coupling strength $\Gamma$ at low temperature $k_BT<\Gamma$, regardless of the ferromagnetic properties of the leads. 
A finite magnetic field lifts the spin degeneracy, and two maxima develop as a function of $\epsilon_0$ with distance $B_z/\Gamma$. Due to a finite temperature and tunnel coupling $\Gamma$, the peaks are not completely separated, i.e. the peak-to-peak distance is less than $B_z/\Gamma=2$ here. The peaks' width is maximal for the antiparallel case and minimal for the parallel setup. Also, the two peaks are symmetric with respect to $\epsilon_0=0$ for all angles $\theta$, since an external magnetic field in $z$-direction does not break the system's particle-hole symmetry at $U=0$.   ISPI data perfectly match the analytical data. 

In Fig.~\ref{fig:BenchLocal} we present (local) observables, occupation and spin-projection expectation values for the same parameters as in Fig.~\ref{fig:BenchCurrent} as a function of $\epsilon_0$.  Again, ISPI data (symbols) are compared to the analytic results (solid lines). We find a doubly occupied dot for large negative gate voltages $\epsilon_0 < -2\Gamma$ and an unoccupied dot for gate voltages $\epsilon_0 > 2\Gamma$. At the particle-hole symmetric point $\epsilon_0=0$ the dot is singly occupied. Differences for the different angles are marginal. Regarding spin expectation values, we find finite $\langle \mathcal{S}_x\rangle$, which is particularly developed for $\theta=0$ and $\theta=\pi/2$ due to resonant tunneling effects, as opposed to lowest-order-in-$\Gamma$ predictions for the quantum-dot spin-valve \cite{Braun_2004}. A small but finite $\langle \mathcal{S}_x\rangle$  for $\theta=\pi$ is addressed to the external magnetic field. We observe symmetric $\langle \mathcal{S}_y\rangle$ curves with respect to $\epsilon_0=0$ for $\theta = \pi$ and $\theta=\pi/2$, whereas $\langle \mathcal{S}_y\rangle=0$ for $\theta=0$, see Eqs.~\eqref{eqs:AnaSpinBz}, where $\langle \mathcal{S}_y\rangle \propto  \sin(\theta/2)$. 
The footprint of the external magnetic field  applied in $z$-direction is reflected in $\langle \mathcal{S}_z\rangle$, where a clear minimum at $\epsilon_0=0$ is visible. Also, the dependence  on the enclosed angle $\theta$ is superposed by the external field and hence marginally detectable.

We note in passing that when applying an external magnetic field in $x$-direction (not shown here), left/right symmetry is broken. The coupling to the leads, i.e.  $\Gamma_{\alpha \pm} = \Gamma_\alpha (1 \pm p)$ for $p\neq 0$, is reflected in a stronger/weaker coupling of majority/minority spins to the dot, which results not only in a splitting of the conductance, but also in different conductance contributions when changing gate voltages. Breaking left/right symmetry is most pronounced for parallel lead magnetizations $\theta=0$ and is reduced for finite $0<\theta<\pi$ and absent for $\theta=\pi$. To sum up our findings for the noninteracting case, we have ensured that ISPI data match perfectly with the analytic solution for all observables and the full range of temperatures, gate and bias voltages as well as lead polarizations and angles.

\subsection{Interaction-induced current asymmetry}
\label{ssec:asymmetry}
\begin{figure}[t]
\centering
\includegraphics[width=\columnwidth]{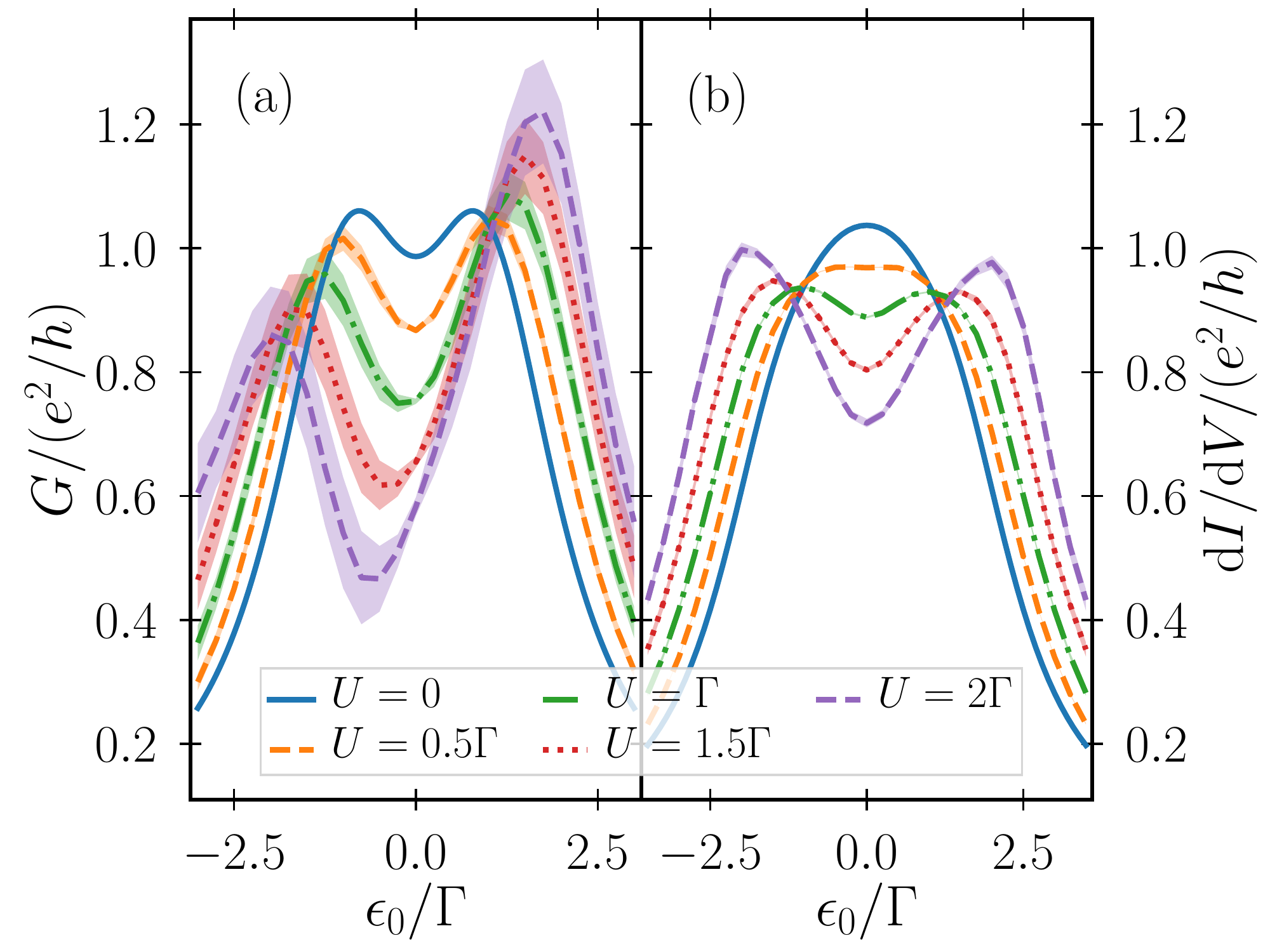}
\caption{(a) Linear and (b) nonlinear differential conductance through the \textit{interacting} spin-valve as a function of gate voltage $\epsilon_0$ for noncollinear lead magnetizations. An external Zeeman field of strength $B_z=2\Gamma$ is applied in $z$-direction. Other parameters are $k_BT=0.2\Gamma$, $\theta = \pi/2$, $p=0.5$, and $eV=0.1\Gamma$ in (a) and $eV=\Gamma$ in (b). Shaded areas are error estimates.}
\label{fig:Geps0Bz}
\end{figure}

We now turn to the case of finite Coulomb interactions, which is relevant for small quantum dots.
 We start out with the settings as in Fig.~\ref{fig:BenchCurrent},  i.e. $k_BT=0.2\Gamma$, $eV=0.1\Gamma$ and $p=0.5$ and perform ISPI calculations for several Coulomb interaction strengths $0\leq U\leq 2\Gamma$ with lead magnetization chosen noncollinearly, $\theta=\pi/2$. For small bias voltages, i.e. the linear conductance regime,  we find the results shown in Fig.~\ref{fig:Geps0Bz}(a). Spin degeneracy is lifted by the Zeeman field and in the absence of interactions $U=0$, see blue solid line, we observe peaks at $\epsilon_0\approx\pm B_z$. Finite Coulomb interaction $U\neq 0$ breaks the spin (particle-hole) symmetry, i.e. $G(\epsilon_0)\neq G(-\epsilon_0)$. Peak positions are renormalized and located at  $\epsilon_0\approx\pm (B_z+U)/2$. Strikingly, the heights are modified according to the strength of the interaction strength. The conductance peak for $\epsilon_0>0$,  attributed to spin $\sigma=\up$-channel is higher whereas its counterpart $\epsilon_0<0$ for $\sigma=\down$ is suppressed. This interaction-induced asymmetry becomes more pronounced for increasing $U$. 
As this effect vanishes for collinear lead magnetizations $\theta=(0,\pi)$ and for $U=0$, we conclude it to be a spin-dependent, \textit{interaction-induced} current asymmetry. Turning to the nonequilibrium setup, i.e. $eV=\Gamma$, see Fig.~\ref{fig:Geps0Bz}(b) we note that the Zeeman splitting is smeared out by the finite bias voltage of comparable order, e.g. the blue solid curve for $U=0$. Nevertheless, finite $U>0$ lifts the spin degeneracy of the energy level and a splitting evolves for interaction strengths $U\geq \Gamma$. Again, with increasing Coulomb interaction strength, the asymmetry of the peak splitting increases. As opposed to the linear-response regime, this effect is less pronounced, probably due to the washing out of quantum effects at large bias voltages. However in the {\it hybrid regime} studied here, nonequilibrium and interaction effects should be treated on the same footing, as is possible with the ISPI method,  in order to obtain a complete physical picture. 

\begin{figure}[t]
\centering
\includegraphics[width=\columnwidth]{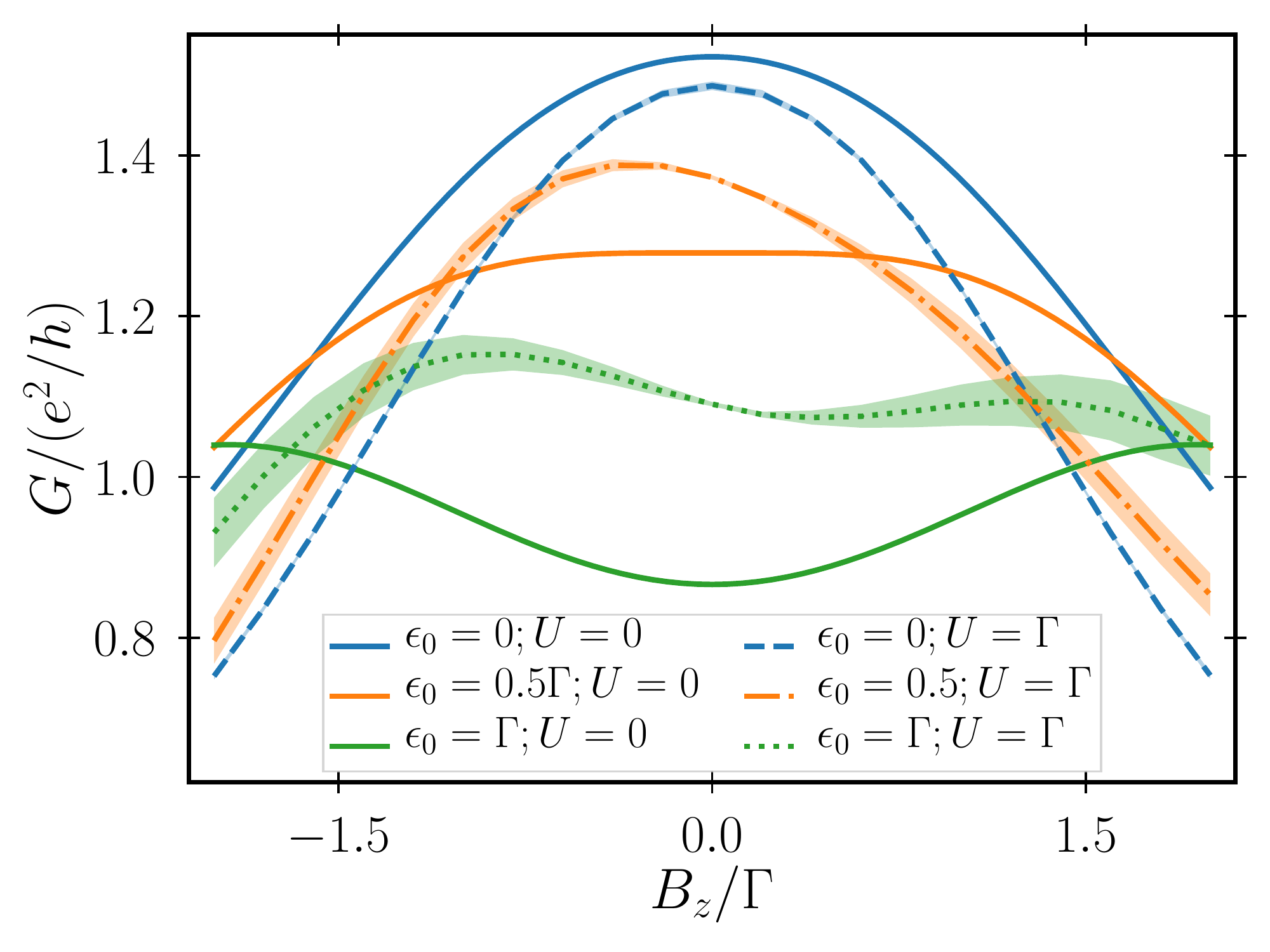}
\caption{Linear conductance as a function of an applied magnetic field $B_z$ for different values of $\epsilon_0$. The noninteracting case (solid lines) is compared to the case of $U=\Gamma$ (shaded areas are error estimate). Other parameters are as in Fig.~\ref{fig:Geps0Bz}(a).}
\label{fig:GBz}
\end{figure}

To further illustrate the appearance of the interaction-induced asymmetry, we show in Fig.~\ref{fig:GBz} the linear conductance $G(B_z)$ for different gate voltages $\epsilon_0$ and Coulomb interactions $U=0$ and $U=\Gamma$. As stated earlier, for $U=0$ all curves are symmetric, which holds also as a function of external magnetic field $B_z$, i.e. left/right symmetry is intact. Finite $U$ instead serves to break this symmetry, and the conductance is sensitive to the sign of the magnetic field, resulting in asymmetric lineshapes in the figure. Note that finite $\epsilon_0\neq 0$ enhances the asymmetry, as the conductance is split by the external magnetic field. 

\begin{figure}
\includegraphics[width=\columnwidth]{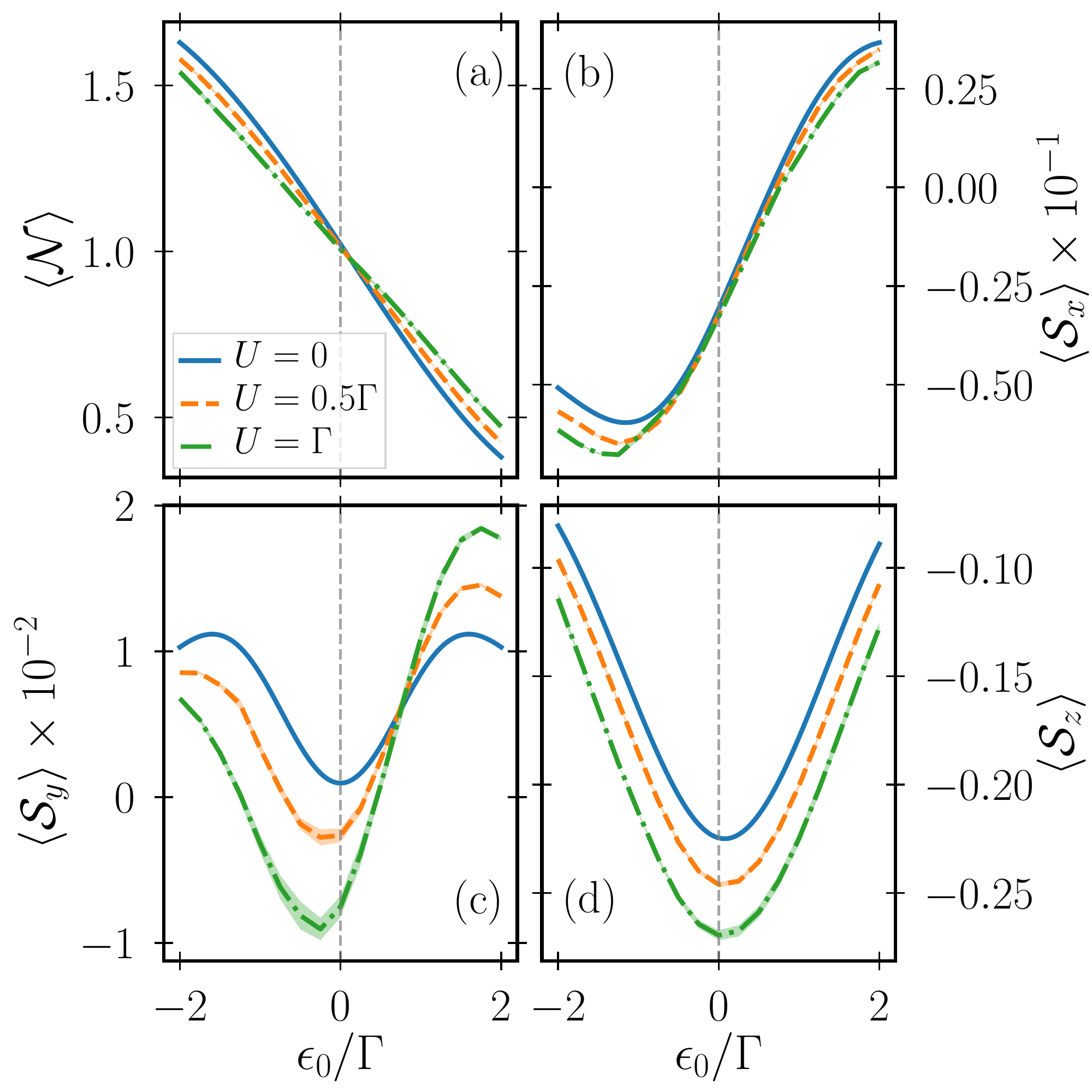}
\caption{Spin expectation values and occupation number in the nonequilibrium regime, see Fig.~\ref{fig:Geps0Bz}(b). ISPI data are given as dashed and dashed-dotted lines, with shaded areas reflecting the corresponding error estimates. Parameters are as in Fig.~\ref{fig:Geps0Bz}.}
\label{fig:OSeps0NonLin}
\end{figure}

In order to complete the picture, we discuss again local observables.  The occupation $\mathcal{N}$ and $\langle \mathcal{S}_{x,y,z}\rangle$ are shown in Figs.~\ref{fig:OSeps0NonLin}(a)-(d) for parameters as in Fig.~\ref{fig:Geps0Bz} and finite bias voltage $eV=\Gamma$. We find that $\expval{\mathcal{S}_x}$ and $\expval{\mathcal{S}_z}$ display a certain asymmetry, also in the noninteracting case, once the bias voltage is increased, which is absent in the linear-response regime (not shown here). Having a finite Coulomb interaction, the exchange field plays a role as well, i.e., an interaction-induced magnetic field which stems from inherent coherences of the system. It has been calculated to first order in the hybridization \cite{Braun_2004,Sothmann_2010_1}.  It is generated by the Coulomb interaction and oriented in direction of the lead magnetizations, of course its particular form depends also strongly on all other system parameters. The impact on $\langle \mathcal{S}_{x,y,z}\rangle$ is visible when increasing Coulomb interaction strength. The $x$- and $z$-components are barely affected qualitatively by the increasing Coulomb interaction, yet they preserve their general behavior and the curves are shifted towards more negative values in the respective expectation value, see panels (b) and (d) in Fig.~\ref{fig:OSeps0NonLin}. For the $y$-component, however, $U\neq 0$ changes the line shape significantly. While for the noninteracting case the $y$-component is suppressed at low bias voltages, see Fig.~\ref{fig:BenchLocal}(c), Coulomb interaction introduces an additional, asymmetric contribution to the $y$-component. With increasing $eV$ the $y$-component is not suppressed anymore, see Fig.~\ref{fig:OSeps0NonLin}(c). 

\subsection{Angular Dependencies}\label{ssec:angle}
\begin{figure}[t]
\centering
\includegraphics[width=\columnwidth]{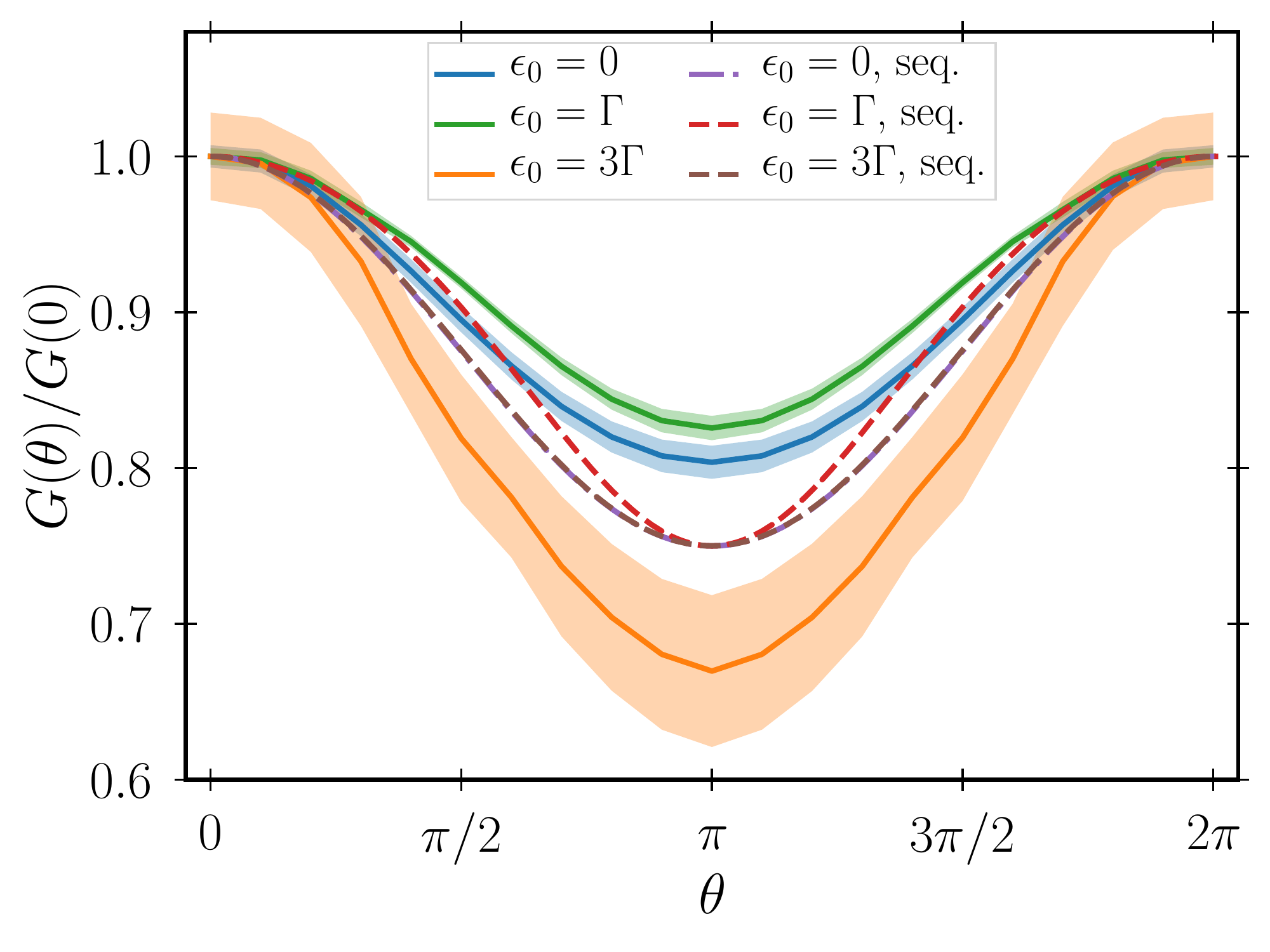}
\caption{Angular dependence of the linear conductance through the quantum-dot spin valve. In the main panel, ISPI data for different $\theta$  are shown (shaded areas are error estimates). For comparison in the inset the results from sequential tunneling (denoted by the shorthand seq.) are given. Parameters are $k_BT=0.2\Gamma$, $U=2\Gamma$, $p=0.5$ and $\vb{B}=0$.}
\label{fig:Itheta}
\end{figure}

In the remainder of our work, we discuss the influence of the angle $\theta$ enclosed by the lead magnetizations on the linear conductance. 
Fig.~\ref{fig:Itheta} shows the linear conductance $G(\theta)/G(\theta=0)$ for parameters as in Fig.~\ref{fig:Geps0Bz} and $U=2\Gamma$ in the main panel. The inset reviews the outcome of the perturbative calculation of Ref.~\onlinecite{Braun_2004}.  Note that $G(\theta)/G(0)$ is closely related to the TMR, see Ref.~\onlinecite{Mundinar_2019}, in fact $\text{TMR} = 1 - G(\pi)/G(0)$. Due to the finite polarization $p$, all datasets have a pronounced minimum at $\theta=\pi$. Taking into account resonant tunneling effects, we observe that the minimal value varies strongly for different gate voltages, $\epsilon_0$, such that $G(\theta)/G(0)|_{\epsilon_0 = \Gamma} < G(\theta)/G(0)|_{\epsilon_0 = 0} < G(\theta)/G(0)|_{\epsilon_0 = 3\Gamma}$. This behavior is consistent with our previous findings \cite{Mundinar_2019}. 

\section{Conclusions}\label{sec:conclusions}
In conclusion, we have studied quantum transport properties of an interacting  quantum-dot spin valve in the resonant tunneling regime by means of the numerically exact ISPI scheme. As an extension to previous works, local observables as dot occupation and spin expectation values have been implemented. The choice of noncollinear magnetization direction of the leads pose no problems and we have performed adequate benchmark checks of the noninteracting system.

Our main finding is an interaction-induced asymmetry in the current profile when changing the gate voltage. An additional Zeeman field together with a small to intermediate Coulomb interaction produces asymmetric current lineshapes, as the spin-symmetry together with left/right-symmetry, is broken. We were able to trace back this asymmetry in the linear and differential conductance to the $y$-component of the spin projection, which is also strongly influenced by Coulomb interactions. The interaction-induced exchange field does highly depend on resonant tunneling effects, which lead to a strong asymmetry effect in the conductance as well as in the spin projections of the quantum dot. This effect is most pronounced at small bias voltages, but does also survive a finite bias voltage that drives the system out of equilibrium, as long as resonant tunneling effects have an impact on the underlying physics.  

Finally, we discussed the dependence of the conductance when varying the angle between the leads' magnetizations. We found significant differences between resonant and sequential tunneling, which are consistent with our earlier predictions of the TMR.

\section{Acknowledgments}
We thank E. Kleinherbers and P. Stegmann for fruitful discussions on spin expectation values and the comparison of our data to the outcome of a lowest-order perturbation theory. 
Financial funding of the Deutsche Forschungsgemeinschaft (DFG, German Research Foundation) under project WE 5733/1-2, KO 1987/5 - 2 and - Projektnummer 278162697 - SFB 1242 is acknowledged.

\appendix
\section{Iterative Summation of Path Integrals}\label{app:ISPI}
In the following, the numerically exact scheme of \textit{iterative summations of path integrals} (ISPI) is summarized. It has been applied to several transport situations for small, interacting quantum dots in recent works\cite{Mundinar_2019, Weiss_2008, Weiss_2013}. The starting point is Eq.~\eqref{eq:discreteGenFunc} in the main text, which poses the problem that summing over all $2^{2N}$ spin configurations is impossible, especially when $N\to \infty$. The ISPI method builds upon the observation that the Green's function $\Delta[0]$, cf. Eq.~\eqref{eq:gom},  of the quantum dot in the presence  of the leads decays exponentially with increasing time differences. This allows us to truncate $\Delta[0]$ for time differences $t-t' = (l-l')\delta_t$ larger than some characteristic correlation time $t_K = K \delta_t$, such that $\Delta[0]=0$ for $\abs{l-l'} \delta_t > t_K$, with $l=1,\ldots, N$. After the truncation we are left with a block tridiagonal matrix $G$ that consist of $4K\times 4K$ blocks $G^{nn'}$, with  $n=1,\dots,N_K=N/K$. For convenience, within our scheme we arranged the appearing HS spins line-wise. It then follows that each block $G^{nn'}$ in its $n$-th row depends on $2K$ distinct HS spins $\vb{s}_n =\left( s^+_{(n-1)K+1}, s^-_{(n-1)K+1}, \ldots,s^+_{nK}, s^-_{nK}\right)$. The determinant of a block tridiagonal matrix can be calculated iteratively \cite{Mundinar_2019, Salkuyeh_2006}, such that  
\begin{equation}
	Z[\eta] = \sum_{\{\vb{s}\}} \prod_{n=1}^{N_K} \det \check G^{nn}
\end{equation}
where $\check G^{nn}$ is defined as
\begin{equation} \label{eq:Gcheck}
	\check G^{nn} = G^{nn} - G^{n,n-1} \left[ \check G^{n-1,n-1} \right]^{-1} G^{n-1,n}
\end{equation}
with the starting condition $\check G^{11} = G^{11}$. 
To keep consistency we approximate $\check G^{nn}$ by
\begin{equation}
	\tilde G^{nn} = G^{nn} - G^{n,n-1} \left[ G^{n-1,n-1} \right]^{-1} G^{n-1,n},
\end{equation}
that is we replace $\left[ \check G^{n-1, n-1}\right]^{-1}$ with $\left[G^{n-1, n-1}\right]^{-1}$ in Eq.~\eqref{eq:Gcheck}. Following this approximation $\tilde G^{nn}$ and consequently its determinant depends on $4K$ HS spins $\vb{s}_n$ and $\vb{s}_{n-1}$. This gives a total of $2^{4K}$ different values of $\det \tilde G^{nn}$, stemming from the different configurations of the HS spins. We arrange these values in a $2^{2K} \times 2^{2K}$ matrix
\begin{equation}
	\Lambda_{n,n-1} = \det \tilde G^{nn}[\vb{s}_n, \vb{s}_{n-1}],
\end{equation}
where each row corresponds to one of the $2^{2K}$ configurations of $\vb{s}_n$ and each column to one of the $2^{2K}$ configurations of $\vb{s}_{n-1}$. We also define a $2^{2K}$ row vector $\bra{1} = (1,1,\ldots,1)$ as well as the $2^{2K}$ starting column vector $\ket{\Lambda_1}$ with 
\begin{equation}
	\ket{\Lambda_1} =\det G^{11}[\vb{s}_1].
\end{equation}
The Keldysh generating functional can then be written in the following compact way
\begin{equation} \label{eq:GenFuncTrans}
	Z[\eta] = \bra{1} \Lambda_{N_K, N_K-1} \ldots \Lambda_{2,1} \ket{\Lambda_1}.
\end{equation}

\section{Spin expectation values}\label{app:Analytical}
Spin expectation values for each projection direction are obtained in the absence of interaction, i.e. $U=0$, analytically. We provide explicit formulae in the presence of a finite magnetic field $B_z$, which we use to benchmark the ISPI data in Sec.~\ref{ssec:benchmark}. In the following we make use of the shorthand notation $c_z^\gamma = [(\omega-\epsilon_0)^2 + \gamma B_z^2 + (\Gamma^2 p^2 \cos(\theta) - \Gamma_{\up} \Gamma_{\down})/2]$ with $\gamma = \pm 1$,
\begin{widetext}
\begin{align}\label{eqs:AnaSpinBz}
	\expval{\mathcal{S}_x} & =\int \frac{\dd{\omega}}{4\pi} \left[
	\frac{2 \Gamma  p \cos \left(\frac{\theta }{2}\right) \left((\omega -\epsilon_0)^2 - B_z^2+\Gamma ^2 \left(p^2 \cos ^2\left(\frac{\theta }{2}\right)-1\right)\right)}{(c_z^-)^2+4 \Gamma ^2 (\omega -\epsilon_0)^2}[f_L(\omega)+f_R(\omega)] \right.\nonumber \\
	&\left. -\frac{4 B_z \Gamma ^2 p \sin \left(\frac{\theta }{2}\right)}{(c_z^-)^2+4 \Gamma ^2 (\omega -\epsilon_0)^2}[f_L(\omega)-f_R(\omega)]\right]\\
	&\nonumber\\	
	\expval{\mathcal{S}_y} &= - \int \frac{\dd{\omega}}{4\pi} \frac{2 \Gamma p \sin(\theta/2) c_z^+ }{(c_z^-)^2 + 4 \Gamma^2 (\omega - \epsilon_0)^2}[f_L(\omega) - f_R(\omega)] \\
	&\nonumber\\
	\expval{\mathcal{S}_z} &= \int \frac{\dd{\omega}}{4\pi} \left[\frac{4 B_z \Gamma (\omega - \epsilon_0)}{(c_z^-)^2 + 4 \Gamma^2 (\omega - \epsilon_0)^2} [f_L(\omega)+f_R(\omega)] + \frac{2 \Gamma^2 p^2 (\omega - \epsilon_0) \sin(\theta)}{(c_z^-)^2 + 4 \Gamma^2 (\omega - \epsilon_0)^2}[f_L(\omega)-f_R(\omega)]\right].
\end{align}
\end{widetext}


\begin{thebibliography}{50}
\bibitem{Parkin_2004} S. S. P. Parkin, Ch. Kaiser, A. Panchula, P. M. Rice, B. Hughes, M. Samant, and S.-H. Yang, \textit{Giant tunnelling magnetoresistance at room temperature with MgO (100) tunnel barriers}, Nat. Mater. \textbf{3}, 862 (2004).
\bibitem{Ikeda_2008} S. Ikeda, J. Hayakawa, Y. Ashizawa, Y. M. Lee, K. Miura, H. Hasegawa, M. Tsunoda, F. Matsukura, and H. Ohno, \textit{Tunnel magnetoresistance of $604\%$ at 300K by suppression of Ta diffusion in CoFeB/MgO/CoFeB pseudo-spin-valves annealed at high temperature}, Appl. Phys. Lett. \textbf{93}, 082508 (2008).
\bibitem{Song_2018} T. Song, X. Cai, M. W.-Y. Tu, X. Zhang, B. Huang, N. P. Wilson, K. L. Seyler, L. Zhu, T. Taniguchi, K. Watanabe, et al., \textit{Giant tunneling magnetoresistance in spin-filter van der Waals heterostructures}, Science \textbf{360}, 1214 (2018).
\bibitem{Gong_2017} C. Gong, L. Li, Z. Li, H. Ji, A. Stern, Y. Xia, T. Cao, W. Bao, C. Wang, Y. Wang, et al., \textit{Discovery of intrinsic ferromagnetism in two-dimensional van der Waals crystals}, Nature \textbf{546}, 265 (2017).
\bibitem{Huang_2017} B. Huang, G. Clark, E. Navarro-Moratalla, D. R. Klein, R. Cheng, K. L. Seyler, D. Zhong, E. Schmidgall, M. A. McGuire, D. H. Cobden, et al., \textit{Layer-dependent ferromagnetism in a van der Waals crystal down to the monolayer limit}, Nature \textbf{546}, 270 (2017).
\bibitem{Hamaya_2007_2} K. Hamaya, M. Kitabatake, K. Shibata, M. Jung, M. Kawamura, K. Hirakawa, T. Machida, T. Taniyama, S. Ishida, and Y. Arakawa, \textit{Electric-field control of tunneling magnetoresistance effect in a Ni / InAs / Ni quantum-dot spin valve}, Appl. Phys. Lett. 91, 022107 (2007).
\bibitem{Sahoo_2005} S. Sahoo, T. Kontos, J. Furer, C. Hoffmann, M. Gräber, A. Cottet, and C. Schönenberger, \textit{Electric field control of spin transport}, Nat. Phys. \textbf{1}, 99 (2005).
\bibitem{Hamaya_2007_1} K. Hamaya, M. Kitabatake, K. Shibata, M. Jung, M. Kawamura, K. Hirakawa, T. Machida, T. Taniyama, S. Ishida, and Y. Arakawa, \textit{Kondo effect in a semiconductor quantum dot coupled to ferromagnetic electrodes}, Appl. Phys. Lett. \textbf{91}, 232105 (2007).
\bibitem{Hauptmann_2008} J. R. Hauptmann, J. Paaske, and P. E. Lindelof, \textit{Electric-field-controlled spin reversal in a quantum dot with ferromagnetic contacts}, Nat. Phys. \textbf{4}, 373 (2008).
\bibitem{Crisan_2016} A. D. Crisan, S. Datta, J. J. Viennot, M.R. Delbecq, A. Cottet, and T. Kontos, \textit{Harnessing spin precession with dissipation}, Nat. Comm. \textbf{7}, 10451 (2016).
\bibitem{Jensen_2005} A. Jensen, J. R. Hauptmann, J. Nyg\r{a}rd, and P. E. Lindelof, \textit{Magnetoresistance in ferromagnetically contacted single-wall carbon nanotubes}, Phys. Rev B \textbf{72}, 035419  (2005).
\bibitem{Braun_2004} M. Braun, J. K\"onig, and J. Martinek, \textit{Theory of transport through quantum-dot spin valves in the weak-coupling regime}, Phys. Rev. B \textbf{70}, 195345 (2004).
\bibitem{Rudzinski_2005} W. Rudzinski, J. Barna\'{s}, R. \'{S}wirkowicz, and M. Wilczy\'{n}ski, \textit{Spin effects in electron tunneling through a quantum dot coupled to noncollinearly polarized ferromagnetic leads}, Phys. Rev. B \textbf{71}, 205307 (2005).
\bibitem{Wenderoth_2016} S. Wenderoth, J. Batge, and R. H\"artle, \textit{Sharp peaks in the conductance of a double quantum dot and a quantum-dot spin valve at high
temperatures: A hierarchical quantum master equation approach}, Phys. Rev. B \textbf{94}, 121303(R) (2016).
\bibitem{Koenig_2003} J. K\"onig and J. Martinek, \textit{Interaction-Driven Spin Precession in Quantum-Dot Spin Valves}, Phys. Rev. Lett. \textbf{90}, 166602 (2003).
\bibitem{Lindebaum_2011} S. Lindebaum and J. K\"onig, \textit{Theory of transport through noncollinear single-electron spin-valve transistors}, Phys. Rev. B \textbf{84}, 235409 (2011).
\bibitem{Lindebaum_2012} S. Lindebaum and J. K\"onig, \textit{Current fluctuations in noncollinear single-electron spin-valve transistors}, Phys. Rev. B  \textbf{86}, 125306 (2012).
\bibitem{Sothmann_2010_1} B. Sothmann, D. Futterer, M. Governale, and J. K\"onig, \textit{Probing the exchange field of a quantum-dot spin valve by a superconducting lead}, Phys. Rev. B \textbf{82}, 094514 (2010).
\bibitem{Hell_2015} M. Hell, B. Sothmann, M. Leijnse, M. R. Wegewijs, and J. K\"onig, \textit{Spin resonance without spin splitting}, Phys. Rev. B \textbf{91}, 195404 (2015). 
\bibitem{Stegmann_2018} P. Stegmann, J. K\"onig, and S. Weiss, \textit{Coherent dynamics in stochastic systems revealed by full counting statistics}, Phys. Rev. B \textbf{98}, 035409 (2018).
\bibitem{Gergs_2018} N. M. Gergs, S. A. Bender, R. A. Duine, and D. Schuricht, {\it Spin Switching via Quantum Dot Spin Valves}, Phys. Rev. Lett. {\bf 120}, 017701 (2018).
\bibitem{Mundinar_2019} S. Mundinar, P. Stegmann, J. K\"onig, S. Weiss, \textit{Iterative path-integral summations for the tunneling magnetoresistance in interacting quantum-dot spin valves}, Phys. Rev. B \textbf{99}, 195457 (2019).
\bibitem{Weymann_2011} I. Weymann, \textit{Finite-temperature spintronic transport through Kondo quantum dots: Numerical renormalization group study}, Phys. Rev. B \textbf{83}, 113306 (2011).
\bibitem{Simon_2007} P. Simon, P. S. Cornaglia, D. Feinberg, and C. A. Balseiro, {\it Kondo effect with noncollinear polarized leads: A numerical renormalization group analysis}, Phys. Rev. B {\bf 75}, 045310 (2007).
\bibitem{Sindel_2007} M. Sindel, L. Borda, J. Martinek, R. Bulla, J. K\"onig, G. Sch\"on, S. Maekawa, and J. von Delft, {\it Kondo quantum dot coupled to ferromagnetic leads: Numerical renormalization group study}, Phys. Rev. B {\bf 76}, 045321 (2007).
\bibitem{Gazza_2006} C. J. Gazza, M. E. Torio, and J. A. Riera, \textit{Quantum dot with ferromagnetic leads: A density-matrix renormalization group study}, Phys. Rev. B \textbf{73}, 193108 (2006).
\bibitem{Swirkowicz_2008} R. \'{S}wirkowicz and M. Wilczy\'{n}ski and J. Barna\'{s}, \textit{ The Kondo effect in quantum dots coupled to ferromagnetic leads with noncollinear magnetizations: effects due to electron-phonon coupling}, Journal of Physics: Condensed Matter \textbf{ 20}, 255219 (2008).
\bibitem{Lichtenstein_2019}  P. Kubiczek, A. Rubtsov, A. I. Lichtenstein, \textit{Exact real-time dynamics of single-impurity Anderson model from a single-spin hybridization-expansion}, SciPost Phys. \textbf{7}, 016 (2019).
\bibitem{Antipov_2017} A. E. Antipov, Q. Dong, J. Kleinhenz, G. Cohen, and E. Gull, {\it Currents and Green's functions of impurities out of equilibrium: Results from inchworm quantum Monte Carlo}, Phys. Rev. B {\bf 95}, 085144 (2017).
\bibitem{Simine_2013} L. Simine and D. Segal, {\it Path-integral simulations with fermionic and bosonic reservoirs: Transport and dissipation in molecular electronic junctions}, J. Chem. Phys. {\bf 138}, 214111 (2013).
\bibitem{Weiss_2013} S. Weiss, R. H\"utzen, D. Becker, J. Eckel, R. Egger, and M. Thorwart, \textit{Iterative path integral summation for nonequilibrium quantum transport}, Phys. Stat. Sol. B \textbf{250}, 2298 (2013).
\bibitem{Weiss_2008} S. Weiss, J. Eckel, M. Thorwart, and R. Egger, \textit{Iterative real-time path integral approach to nonequilibrium quantum transport}, Phys. Rev. B \textbf{77}, 195316 (2008); Phys. Rev. B \textbf{79}, 249901(E) (2009).
\bibitem{Weiss_2015} S. Weiss, J. Br\"uggemann, and M. Thorwart, \textit{Spin vibronics in interacting nonmagnetic molecular nanojunctions}, Phys. Rev. B \textbf{92}, 045431 (2015).
\bibitem{Kamenev} A. Kamenev, \textit{Field theory of non-equilibrium systems}, (Cambridge University Press, Cambridge, 2011)
\bibitem{Negele_Orland} J. W. Negele and H. Orland, \textit{Quantum Many-Particle Systems}, (Addison-Wesley, Redwood City, CA, 1988).
\bibitem{Hubbard_1959} J. Hubbard, \textit{Calculation of Partition Functions}, Phys. Rev. Lett. \textbf{3}, 77 (1959).
\bibitem{Hirsch_1983} J. E. Hirsch, \textit{Discrete Hubbard-Stratonovich transformation for fermion lattice models}, Phys. Rev. B \textbf{28}, 4059 (1983).
\bibitem{Jauho_1994} A. P. Jauho, N. S. Wingreen, and Y. Meir, \textit{Time-dependent transport in interacting and noninteracting resonant-tunneling systems}, Phys. Rev. B \textbf{50}, 5528 (1994).
\bibitem{Salkuyeh_2006} D. K. Salkuyeh, \textit{Comments on "A note on a three term recurrence for a tridiagonal matrix"}, Appl. Math. Comp. \textbf{176}, 442 (2006).

\end{thebibliography}
\end{document}